\documentclass[reprint, amsmath, amssymb, showkeys, superscriptaddress,
aip]{revtex4-1}

\usepackage{array}
\usepackage{gensymb}
\usepackage{graphicx}

\newcommand{\sevenabreast}{0.064\textwidth}

\newcommand{\affiliationsoton}{Faculty of Engineering and the Environment,
  University of Southampton, Southampton, SO16 7QF, United Kingdom.}

\begin{document}

\title{Skyrmions in thin films with easy-plane magnetocrystalline anisotropy}

\author{Mark~Vousden}
\affiliation{\affiliationsoton}
\author{Maximilian~Albert}
\affiliation{\affiliationsoton}
\author{Marijan~Beg}
\affiliation{\affiliationsoton}
\author{Marc-Antonio~Bisotti}
\affiliation{\affiliationsoton}
\author{Rebecca~Carey}
\affiliation{\affiliationsoton}
\author{Dmitri~Chernyshenko}
\affiliation{\affiliationsoton}
\author{David~Cort\'es-Ortu\~no}
\affiliation{\affiliationsoton}
\author{Weiwei~Wang}
\affiliation{Department of Physics, Ningbo University, Ningbo 315211, China}
\author{Ondrej~Hovorka}
\affiliation{\affiliationsoton}
\author{Christopher~H.~Marrows}
\affiliation{School of Physics \& Astronomy, University of Leeds, Leeds LS2
  9JT, United Kingdom.}
\author{Hans~Fangohr}
\affiliation{\affiliationsoton}

\date{\today}

\begin{abstract}
    We demonstrate that chiral skyrmionic magnetization configurations can be
    found as the minimum energy state in B20 thin film materials with
    easy-plane magnetocrystalline anisotropy with an applied magnetic field
    perpendicular to the film plane.
    Our observations contradict results from prior analytical work, but are
    compatible with recent experimental investigations.
    The size of the observed skyrmions increases with the easy-plane
    magnetocrystalline anisotropy.
    We use a full micromagnetic model including demagnetization and a
    three-dimensional geometry to find local energy minimum (metastable)
    magnetization configurations using numerical damped time integration. We
    explore the phase space of the system and start simulations from a variety
    of initial magnetization configurations to present a systematic overview of
    anisotropy and magnetic field parameters for which skyrmions are metastable
    and global energy minimum (stable) states.
\end{abstract}


\maketitle

Skyrmions are topological defects\cite{skyrme61} that can be observed in the
magnetization configuration of materials that lack inversion
symmetry,\cite{bogdanov94} either due to a non-centrosymmetric crystal
lattice,\cite{dzyaloshinskii58,moriya60} or at interfaces between different
materials.\cite{crepieux98} This lack of inversion symmetry results in a chiral
interaction known as the Dzyaloshinskii-Moriya (DM)
interaction.\cite{dzyaloshinskii58,moriya60} The DM interaction results in a
rich variety of chiral magnetization configurations, including helical,
conical, and skyrmionic magnetization configurations. Skyrmionic configurations
were predicted\cite{rossler06} and later observed in helimagnetic materials,
\cite{muhlbauer09,yu10,yu11,seki12} and materials with an interfacial DM
interaction.\cite{woo15,moreau16,jiang15,boulle16,stebliy15}

Skyrmions demonstrate potential for applications in data storage and processing
devices. Skyrmions have been observed with diameters of the order of atom
spacings in mono-atomic Fe layers,\cite{heinze11} which is significantly
smaller than the magnetic domains proposed for the racetrack memory
design.\cite{parkin08} This results in a greater storage density. The movement
of skyrmions has also been demonstrated\cite{jonietz10,yu12} using
spin-polarized current densities of the order $10^6\,\mathrm{Am}^{-2}$, which
is orders of magnitude less than what is required to move magnetic domain
walls.\cite{boulle11,parkin08} These observations demonstrate potential for
skyrmion-based racetrack memory technology \cite{fert13} and other data storage
and processing devices.\cite{zhang15}

Certain material restrictions need to be overcome before skyrmions can be used
in such technologies. While skyrmions can be stabilized, they are only stable
in a limited region of the parameter space defined by an applied magnetic field
and the temperature. This region is narrow in bulk materials,\cite{muhlbauer09}
larger in thin film materials,\cite{yu11} and further stabilized in laterally
confined geometries \cite{beg15} and materials with pinning
defects.\cite{lin13} Analytical analysis of helimagnetic thin film material
models find that skyrmion lattice states are ground states in helimagnetic thin
films with an applied magnetic field only in systems with easy-axis
magnetocrystalline anisotropy,\cite{wilson14,bogdanov94} where the easy axis
and the applied field are perpendicular to the plane of the film. However,
simulated annealing methods find that skyrmions can be the ground state in
two-dimensional helimagnetic thin films with easy-plane anisotropy.\cite{lin15}
Skyrmions have also been identified in two-dimensional surface-inversion
breaking systems with easy-plane anisotropy and Rashba spin-orbit
coupling.\cite{rowland16} Furthermore, experimental studies of easy-plane
helimagnetic thin films identify an additional contribution to the Hall
resistivity beyond the ordinary and anomalous contributions.
\cite{huang12,li13,porter13,meynell14,porter14,yokouchi14} This may be
interpreted as the topological Hall effect, which arises through real space
Berry phase effects,\cite{bruno04,binz08} and is an indication of potential
skyrmion presence. Skyrmions have been directly observed with Lorentz
transmission electron microscopy (LTEM) in easy-plane MnSi with the field
applied along all principal crystallographic directions.\cite{yu15}

In this letter, prior analyses are extended by considering a three-dimensional
thin film with demagnetization to determine whether or not skyrmions are stable
in thin films with easy-plane anisotropy. For a magnetization configuration to
be stable, it must be a configuration with the lowest possible
energy. Variational techniques have been used to minimize the energy in
simplified model systems analytically. \cite{bogdanov99} Here we use numerical
simulation methods to solve a more complete model system.

We consider a cuboidal simulation cell representing a thin film of
Fe$_{0.7}$Co$_{0.3}$Si. The cell has lateral dimensions $L_x$ and $L_y$, and
finite thickness $L_z\ll L_x,L_y$, where $x$, $y$, and $z$ are Cartesian axes
with origin at the center of the geometry. $L_x$ is equal to the helical
period, and $L_y=L_x\sqrt{3}$ to support hexagonal skyrmion lattice
magnetization configurations in the simulation cell. Periodic boundary
conditions are imposed on the Heisenberg and DM exchange interactions in the
lateral directions. The macrogeometry approach\cite{fangohr09} is used to model
periodicity of the demagnetizing field, with a disc macrogeometry of radius
equal to 26 times the helical period and thickness $L_z=5\,\mathrm{nm}$. The
boundary conditions pose a mathematically different problem from analytical
work conducted previously, which considers rotationally symmetric
magnetization.\cite{bogdanov94} Our method allows arbitrary magnetization
configurations that do not satisfy this symmetry, such as helical states.

The standard numerical micromagnetics approach of solving the
Landau-Lifshitz-Gilbert (LLG) equation as an initial value problem is employed
here. Unlike previous work, demagnetization effects are incorporated in this
energy model since demagnetization is known to affect the stability of
skyrmions.\cite{beg15} A three-dimensional simulation domain with finite
thickness is used because magnetization variation in the thickness direction
can stabilize skyrmionic configurations.\cite{rybakov13} These extensions
differentiate this study from previous works. \cite{bogdanov94,wilson14,lin15}

The micromagnetic representation of the system energy is modelled here as
\begin{equation}
W(\mathbf{m})=\int_V\left(w_\mathrm{e}+w_\mathrm{dmi}+w_\mathrm{z}+w_\mathrm{a}
+w_\mathrm{d}\right)\,\mathop{\mathrm{d}V},
\label{eq:energy}
\end{equation}
where $V$ is a cuboid region of volume $L_x\times L_y\times L_z$, and
$\mathbf{m}(x,y,z)=\mathbf{M}(x,y,z)/M_\mathrm{S}$ is the magnetization vector
field normalized by saturation magnetization $M_\mathrm{S}$ such that
$\mathbf{|m|}=1$. The terms $w_\mathrm{e}=A\left(\nabla\mathbf{m}\right)^2$ and
$w_\mathrm{dmi}=D\mathbf{m}\cdot\left(\nabla\times\mathbf{m}\right)$ are the
energy density contributions from Heisenberg and DM exchange interactions
respectively. The term
$w_\mathrm{z}=-\mu_0M_\mathrm{S}\mathbf{H}\cdot\mathbf{m}$ is the Zeeman energy
density contribution from the applied magnetic field. The term
$w_\mathrm{a}=K_1(1-\left(\mathbf{m}\cdot\hat{\mathbf{z}})^2\right)$ is the
energy density contribution from the magnetocrystalline anisotropy. This
anisotropy is easy-axis when \hbox{$K_1>0$} and is easy-plane when
\hbox{$K_1<0$}. The term
$w_\mathrm{d}=-\mu_0M_\mathrm{S}(\mathbf{H}_\mathrm{d}\cdot\mathbf{m})/2$ is
the energy density contribution from demagnetization, where the demagnetizing
field $\mathbf{H}_\mathrm{d}$ is calculated using the Fredkin-Koehler finite
element method-boundary element method (FEMBEM).\cite{fredkin90}

The energy model uses material parameters from experiments on
Fe$_{0.7}$Co$_{0.3}$Si\cite{sinha14,porter13} as follows: the symmetric
exchange coefficient \hbox{$A=4.0\times10^{-13}\,\mathrm{Jm^{-1}}$}, the DM
exchange coefficient \hbox{$D=2.7\times10^{-4}\,\mathrm{Jm^{-2}}$}, the
magnetocrystalline anisotropy coefficient
\hbox{$K_1=-3.0\times10^4\,\mathrm{Jm^{-3}}$}, and the saturation magnetization
\hbox{$M_\mathrm{S}=9.5\times10^4\,\mathrm{Am^{-1}}$}. To obtain a systematic
data set and understanding, the magnetocrystalline anisotropy $K_1$ is varied
in the range \hbox{$[-0.5K_0,1.25K_0]$}, where
\hbox{$K_0=D^2/A=1.8\times10^5\,\mathrm{Jm^{-3}}$}. This range contains the
anisotropy value \hbox{$K_1=-0.16K_0$} calculated for
Fe$_{0.7}$Co$_{0.3}$Si. The applied field parallel to the $z$ direction
$|\mathbf{H}|$, is varied in the range \hbox{$[0,15.4M_\mathrm{S}]$}.

\begin{figure}[t]
    \centering
    \includegraphics{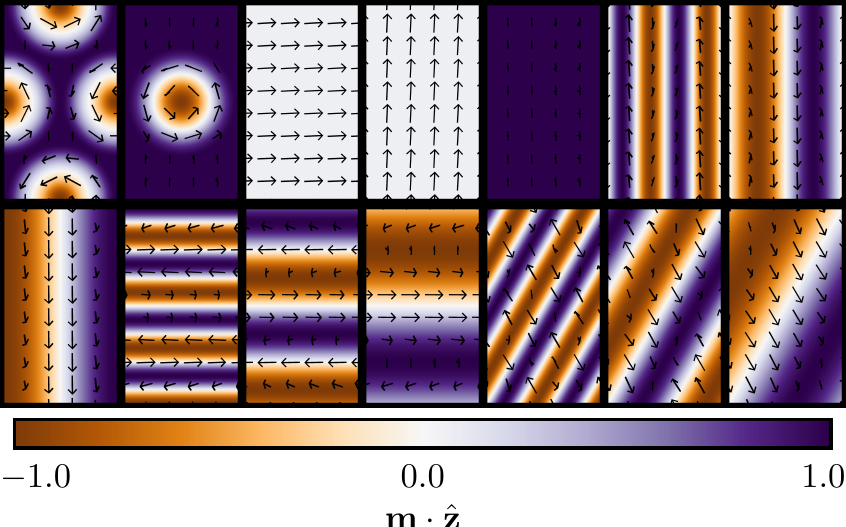}
    \caption{Initial magnetization configurations to be relaxed using damped
      Landau-Lifshitz-Gilbert (LLG) dynamics. From the top, left to right,
      these are hexagonal and rectangular skyrmion lattice states, canted
      uniform states parallel to the $x$, $y$, and $z$ directions, and a
      variety of helical states. The period of the helices is varied to support
      metastable states with different helical periods.}
    \label{fig:initial_states}
\end{figure}

\begin{figure}[t]
    \centering
    \includegraphics{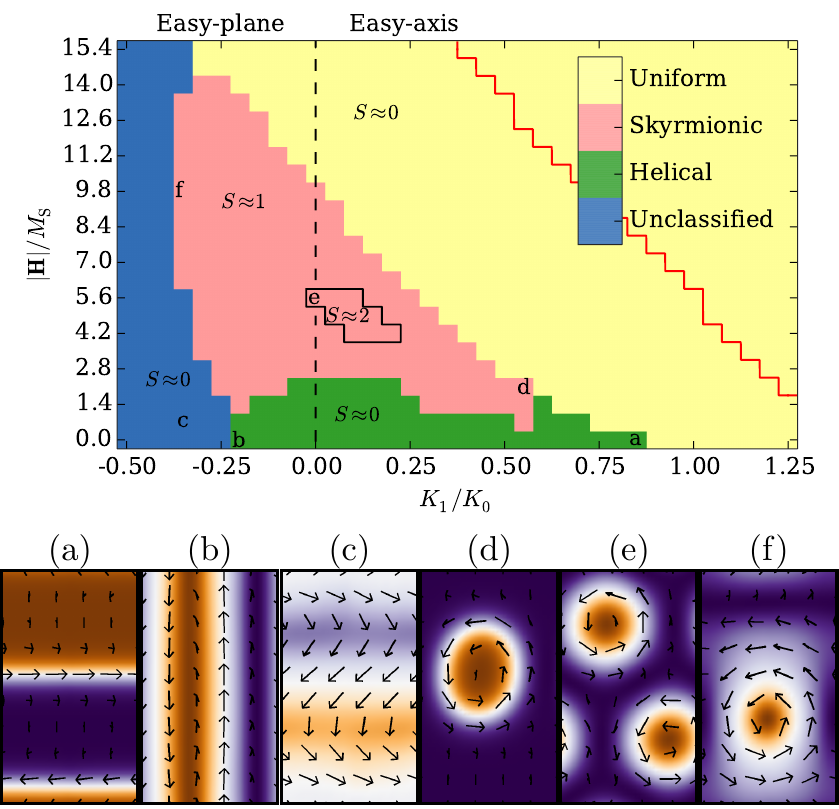}
    \caption{Top: Phase diagram showing ground state variation with normalized
      anisotropy $K_1/K_0$ and applied field magnitude
      $|\mathbf{H}|/M_\mathrm{S}$. Skyrmions are ground states in both
      easy-plane and easy-axis anisotropy regions. Parameters below the solid
      line in the uniform region support metastable skyrmionic states. Bottom:
      Selected magnetization configurations.}
    \label{fig:phase_diagram}
\end{figure}

Multiple initial magnetization configurations are used to determine the minimum
energy state of the micromagnetic system for each combination of $|\mathbf{H}|$
and $K_1$ values. These initial states are shown in
Fig.~\ref{fig:initial_states}. A finite element method-based simulator has been
used to solve the damped LLG equation as an initial value problem for the
aforementioned system. The simulator is similar to the software
Nmag\cite{fischbacher07} and uses the FEniCS \cite{logg10} finite element
framework. The edge lengths of the tetrahedral elements of the finite element
mesh do not exceed one nanometer, which is chosen to be smaller than both the
Bloch parameter \hbox{$\sqrt{A/|K_1|}>1.3\,\mathrm{nm}$} and the exchange
length \hbox{$\sqrt{2A/\mu_0M_\mathrm{S}^2}=8.4\,\mathrm{nm}$} to correctly
resolve micromagnetic behavior. The same finite element mesh is used to
discretize the magnetization domain for all results reported here. When the
magnetization precesses more slowly than \hbox{$1\,\degree\mathrm{ns^{-1}}$}
everywhere, it is considered relaxed. All initial configurations shown in
Fig.~\ref{fig:initial_states} are relaxed independently for each pair of
$|\mathbf{H}|$ and $K_1$ values. The energies of all relaxed configurations at
the same $|\mathbf{H}|$ and $K_1$ are compared, and the configuration with the
lowest energy out of all these states is classified as the ground state. A
selection of these states are shown in Fig.~\ref{fig:phase_diagram} (a) to
(f). Table~\ref{tbl:phase_diagram} shows the normalized $|\mathbf{H}|$ and
$K_1$ values for each of the six configurations (a) to (f).

Fig.~\ref{fig:phase_diagram} (top) shows a phase diagram which groups relaxed
configurations into uniform, skyrmionic, helical, and unclassified
configurations. This phase diagram shows the obtained ground state
configuration for each point in the applied field and anisotropy parameter
space. The locations of the ground states shown in Fig.~\ref{fig:phase_diagram}
(a) to (f) are indicated in the phase diagram.

States are considered as uniform if $\mathbf{m}\cdot\bar{\mathbf{m}}>0.85$
everywhere, where $\bar{\mathbf{m}}$ is the spatially-averaged magnetization
direction. States are considered skyrmionic if the skyrmion number
$S(\mathbf{m})>0.5$, where
\begin{equation}
S(\mathbf{m})=\frac{1}{4\pi}\int_T\mathbf{m}\cdot\left(\frac{\partial
\mathbf{m}}{\partial x}\times\frac{\partial\mathbf{m}}{\partial y}\right)
\mathop{\mathrm{d}x}\mathop{\mathrm{d}y},
\label{eq:skyrmion_number}
\end{equation}
and $T$ is the surface at $z=0$ contained by $V$. States are considered helical
if the magnetization configuration contains a full rotation along a single
direction in the thin film. Magnetization configurations that do not satisfy
any of these conditions are unclassified, but are still considered when
energies of relaxed states are compared.

\begin{table}[b]
    \caption{Normalized anisotropy and applied field values for the selected
      magnetization configurations in Fig.~\ref{fig:phase_diagram} (a) to (f).}
    \label{tbl:phase_diagram}
    \begin{ruledtabular}
    \begin{tabular}{*{7}{>{\centering\arraybackslash}p{\sevenabreast}}}
        &(a)&(b)&(c)&(d)&(e)&(f)\\ \hline
        $K_1/K_0$&$0.85$&$-0.20$&$-0.35$&$0.55$&$0.00$&$-0.35$\\
        $|\mathbf{H}|/M_\mathrm{S}$&$0.0$&$0.0$&$0.7$&$2.1$&$5.6$&$9.8$\\
    \end{tabular}
    \end{ruledtabular}
\end{table}

The helical configuration is found for anisotropy values $K_1$ in the range
$[-0.2K_0,0.85K_0]$, and weak applied fields, in agreement with previous
predictions \cite{bak80} and observations.\cite{yu11} The uniform state is
observed when the applied field and easy-axis anisotropy dominate, and these
states have magnetization aligned with the out-of-plane direction ($z$). This
is expected, since the energy contributions from the applied field and the
easy-axis magnetocrystalline anisotropy are minimized in this case. The
magnetization configurations in the unclassified region, exemplified by
Fig.~\ref{fig:phase_diagram}~(c), are driven by the easy-plane anisotropy
contribution, which causes the magnetization to orient mostly within the plane
of the thin-film.

Skyrmion lattice states (Fig.~\ref{fig:phase_diagram} (d) to (f)) are minimum
energy states for both positive (easy-axis) and negative (easy-plane) values of
the magnetocrystalline anisotropy coefficient. Analytical work suggests that
skyrmion lattice states are minimum energy states only for positive anisotropy
values $K_1$ in the range $[0,0.48K_0]$,\cite{wilson14} which is in contrast to
the wider $[-0.35K_0,0.55K_0]$ range observed in this work, which includes both
positive and negative anisotropy values. There are differences between our work
and the model used in Ref.~\onlinecite{wilson14} that we believe causes this
discrepancy. Firstly, unlike the analytical work, our model includes
demagnetization energy, which is known to change skyrmion
energetics.\cite{rybakov13, beg15} Secondly, their work identifies skyrmion
lattice states as local energy minima and the cone state as the ground state in
easy-plane anisotropy systems. In our work, the thinner films suppress conical
states,\cite{yu11} resulting in skyrmion lattice states having the lowest
energy. Finally, the chiral twist in the magnetization at the top and bottom of
the film, that are known to contribute to the skyrmion stability,
\cite{leonov15, beg15} are accounted for in our model, unlike the
two-dimensional analytical model where the skyrmion magnetization does not vary
in the thickness direction.

The parameter sets within the solid line in the skyrmion region of
Fig.~\ref{fig:phase_diagram} exhibit a hexagonal skyrmion lattice as the
minimum energy configuration, as shown in Fig.~\ref{fig:phase_diagram}~(e). The
magnetization in this simulation cell has a skyrmion number $S\approx2$. In the
remaining skyrmionic region, one skyrmion per simulation cell is found (as in
Fig.~\ref{fig:phase_diagram}~(d) and (f)), which corresponds to a rectangular
skyrmion lattice ($S\approx1$). The rectangular skyrmion lattice has a lower
energy than the hexagonal lattice over most of the parameter space. This occurs
because the skyrmion spacing that minimizes the energy changes with the
anisotropy, meaning that the hexagonal skyrmions are frustrated in the
simulation cell. These two skyrmion lattice configurations compete.\cite{lin15}

\begin{figure}[t]
    \centering
    \includegraphics{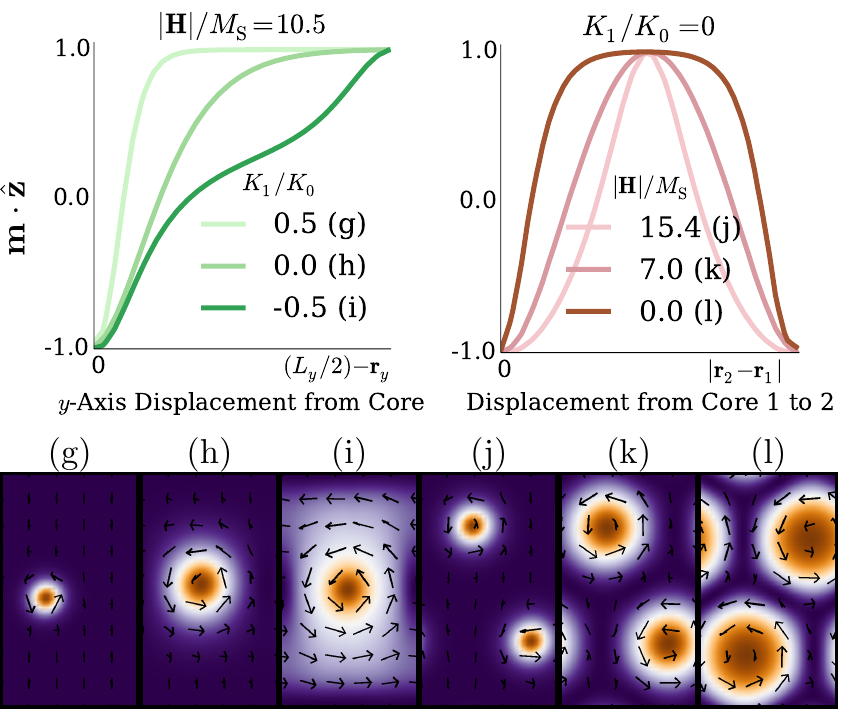}
    \caption{Top: Metastable skyrmion profiles varying with anisotropy in the
      rectangular lattice (left), and applied field magnitude in the hexagonal
      lattice (right), where $\mathbf{r}$ is the coordinate of the skyrmion
      core. Bottom: Magnetization configurations of the profiles shown above.}
    \label{fig:skyrmion_size}
\end{figure}

\begin{table}[b]
    \caption{Normalized anisotropy and applied field values for the selected
      magnetization configurations in Fig.~\ref{fig:skyrmion_size} (g) to (l).}
    \label{tbl:parametersforprofilesnapshots}
    \begin{ruledtabular}
    \begin{tabular}{*{7}{>{\centering\arraybackslash}p{\sevenabreast}}}
        &(g)&(h)&(i)&(j)&(k)&(l)\\ \hline
        $K_1/K_0$&$0.50$&$0.00$&$-0.50$&$0.00$&$0.00$&$0.00$ \\
        $|\mathbf{H}|/M_\mathrm{S}$&$10.5$&$10.5$&$10.5$&$15.4$&$7.0$&$0.0$ \\
    \end{tabular}
    \end{ruledtabular}
\end{table}

Parameter sets below the solid line in the uniform region of
Fig.~\ref{fig:phase_diagram} support local energy minimum (metastable) skyrmion
states, which encompasses most of the parameter
space. Fig.~\ref{fig:skyrmion_size} shows how metastable skyrmion size changes
with anisotropy (snapshots (g), (h) and (i)) and applied field magnitude
(snapshots (j), (k) and (l)) at $z=0$, as parameterized by
Table~\ref{tbl:parametersforprofilesnapshots}. Skyrmion size increases with
decreasing magnetic field magnitude and increasing easy-plane anisotropy
because the magnetization component in the plane is more favorable
energetically, which increases the length over which the skyrmion
twists. Fig.~\ref{fig:skyrmion_size}~(i) shows that skyrmions expand to fill
the geometry that contains them when easy-plane anisotropy is strong
enough. This causes the skyrmion to stretch into two separate
objects.\cite{lin15}

To summarize, skyrmions are minimum energy states in magnetic systems with
easy-plane magnetocrystalline anisotropy and demagnetizing effects under
micromagnetic simulation. This conclusion is in counterpoint to findings with
analytical models,\cite{wilson14, bogdanov94} but agrees with numerical work
that does not consider demagnetization or thickness effects.\cite{lin15}
Systems with weaker easy-plane anisotropy result in smaller skyrmions.
Skyrmion lattice states are also metastable in a wide range of anisotropy
values. The skyrmion presence as a minimum energy state is consistent with LTEM
observations,\cite{yu15} and supports the interpretation of Hall signals in
easy-plane thin films as the topological Hall
effect.\cite{huang12,porter13,li13,porter14,meynell14,yokouchi14} This
encourages research into a broader range of materials for skyrmion physics and
spintronic applications,\cite{tokunaga15} as it may enhance the stability of
skyrmions in systems where they are already favorable, such as confined
geometry materials, and materials with pinning defects. \cite{lin13}

We acknowledge financial support from EPSRC's DTC grant EP/G03690X/1 and
standard research grant EP/J007110/1.

\bibliography{paper}
\end{document}